\documentclass[%
 reprint,
 onecolumn,
groupedaddress,
 amsmath,amssymb,
 aps,
]{revtex4-2}

\draft

\usepackage{subfigure}
\usepackage{graphicx}
\usepackage{dcolumn}
\usepackage{bm}
\usepackage{xcolor}
\usepackage{soul}

\begin{document}

\preprint{APS/123-QED}

\title{Interaction between swarming active matter and flow: the impact on Lagrangian coherent structures}

\author{Xinyu Si}
\author{Lei Fang}
 \email{lei.fang@pitt.edu}
\affiliation{
Department of Civil and Environmental Engineering,\\
University of Pittsburgh, Pittsburgh, Pennsylvania 15261, USA
}

\begin{abstract}

In recent years, research topics concerning active matter have attracted interest from diverse communities. It has been suggested that active matter - as represented by organisms such as zooplankton - has great potential in ocean mixing due to its intrinsic mobility and the sheer amount of biomass. 
However, prior investigations have predominantly overlooked the influence of external background flow, despite the ubiquity of flow driven by various sources in nature. The interaction between active matter and external flow structures has long been neglected. Here, we conducted experiments using a typical centimeter swimmer, \textit{A. salina}, and an electromagnetically driven quasi-two-dimensional flow to study the interaction between active matter and flow. We focused on the impact of swarming active matter on hyperbolic Lagrangian coherent structures (LCSs) that mark the most straining regions in the flow. There is one decade of scale separation between active matter agents and the length scale of LCSs. We illustrated that the impact of active matter on LCSs was much more significant compared to localized random noise with similar energy input. In addition, we revealed that the perturbation generated by active matter could couple with the background flow and further deform the LCSs. In addition to the impact on the most straining hyperbolic regions of flow, we also revealed that the rotational elliptical region of the flow was much more susceptible to active matter perturbation. We further described how the influence of active matter changed with their number densities and background flow intensities. We revealed that the LCSs could be decently altered even at a small number density of active matter. Through this work, we aim to provide valuable insights and draw attention to the problem regarding the interaction between active matter and external flow structures.

\end{abstract}

\maketitle

\section{Introduction}\label{Introduction}
Active matter defines a broad spectrum of subjects that are characterized by the capability of transducing free energy into systematic movements \cite{ramaswamy2010mechanics}. Under this broad definition, active matter encompasses both living matter and nonliving imitations, ranging from micrometer-scale bacteria to dekameter-scale mammals \cite{marchetti2013hydrodynamics,brambilla2013swarm}. This comprehensive inclusion of research subjects hence attracted the attention of diverse communities with varying perspectives.

One major perspective focuses on the collective behaviors and self-organization of active matter. It is ubiquitous in natural systems that huge amounts of living matter, such as collections of bacteria and fish schools, exhibit coherent motions and long-range ordering. The analogy between flocking active matter and some condensed matter systems like ferromagnets and liquid crystals \cite{toner2005hydrodynamics} has captivated the interest of and spawned extensive investigation within the communities of condensed matter physics and statistical physics. The viewpoint is to regard the collection of active matter as some materials so that the collective behaviors of active matter can be characterized and understood through the adaptation and modification of suitable condensed matter physics theories \cite{van2019mechanical} and hydrodynamics theories \cite{ramaswamy2010mechanics,bain2019dynamic}. Research topics from this perspective are vast and diverse. Some of the main research topics include emergent structures, symmetries, phase transitions, and rheological properties \cite{marchetti2013hydrodynamics}.

This paper, however, shifts the viewpoint from active matter behaviors to the flow environment they are living in. 
Active agents consume energy in order to move around and exert mechanical forces on the flow around them \cite{ramaswamy2010mechanics,marchetti2013hydrodynamics}. 
Especially for aquatic creatures, the enormous range of body size covered by various aquatic animals and the relatively higher viscosity of liquid than that of air result in the expansive Reynolds numbers for active-matter-induced flow, from Stokes flow to fully developed turbulence \cite{huntley2004influence}. The diverse moving patterns of various species make the classification and characterization of swimmer-generated flows even more challenging and intriguing. There are many investigations conducted just focusing on the locomotion and the detailed description of flows induced by various moving patterns of different species under different regimes of Reynolds numbers \cite{lauga2016bacterial, guasto2012fluid, lauder2007fish}. 

Another considerable portion of investigations focuses on the chaotic flows at very low Reynolds numbers, where inertia effects are negligible. These chaotic flows are also known as active turbulence \cite{alert2022active}. The constituents of the active matter for active turbulence are mostly micro-swimmers, such as bacteria \cite{wu2000particle,peng2021imaging} and tissue cells \cite{lin2021energetics}. The key difference between active turbulence and inertial turbulence is that the active matter is not treated as external forces but rather a component in a continuum description of the flow system. Consequently, the active turbulence is considered as autonomous and self-organized. The investigations of active turbulence are closely related to the research with respect to the collective movements of active matter. Attentions are paid to the characterization and classification of the active turbulence under different ordering phases of flocking active matter \cite{alert2022active}. 

The investigation on biologically generated turbulence with higher Reynolds numbers were primarily conducted by the oceanographic community focusing on the mixing effect of aquatic animals.
Early stage investigations mainly focused on a bulk estimation of the injected energy by marine animals and hence evaluating their potential on mixing the ocean \cite{huntley2004influence, dewar2006does, kunze2006observations}. Contrary opinions have been raised \cite{visser2007biomixing, kunze2019biologically} regarding the mixing effect for low-Reynolds-number swimmers due to their low mixing efficiency. However, mesoscale animals are at least still promising in ocean mixing due to their massive total amounts and the intermediate Reynolds numbers \cite{katija2012biogenic}. Additionally, in recent years, it was found that collective motion of centimeter-scale aquatic organisms could form aggregate-scale eddies that could effectively modify the stratification of water columns at much larger length scales \cite{houghton2018vertically, ouillon2020active}.

Even though much research has been performed on active-matter-induced flow, almost all previous works have considered the active matter as energy sources independently without considering the background flow environment. However, natural water bodies like oceans and lakes are driven by various of energy sources and hence the living environment of most aquatic creatures are almost always flowing. The interaction between biologically generated flow and the background flow environment can be a significant factor on influencing the flow environment but has long been overlooked. In one of our recent research, we performed investigations on the interaction between a biologically generated jet behind a centimeter-scale swimmer and a background shear flow. We found that both the intensity and direction of the biologically generated turbulent energy cascade depended on the geometric configuration between the biologically generated agitation and the background shear. As a result, the background shear could be either intensified or attenuated depending on different tensor geometries \cite{Si2023Biologically}.

Here, we continue on the investigation about the interaction between active-matter-induced agitations and background flow fields. In this work, we incorporated a typical centimeter-scale swimmer \textit{A. salina}, commonly known as brine shrimp, under swarming movement into an electromagnetically drive quasi-two-dimensional (quasi-2D) cellular flow. By ``swarming", we mean the polarization \cite{attanasi2014information,ling2019collective} of the group is, on average, zero, and the velocity of the agents is randomly aligned with each other. We focus on the impact of the swarming active matter on Lagrangian coherent structures (LCSs). LCSs characterize the flow in a simpler lower-order way. They form the skeletons of the flow, where the strongest straining occurs. Hyperbolic LCSs are considered as good detection of transport barriers. Therefore, studying the influence on LCSs is also a good way to diagnose the impact on transport and mixing properties of the flow.

We start with providing experimental details in \S \ref{2Dflow} and \S \ref{Sec_ActiveMatter}. Section \ref{2Dflow} describes the experimental methods used to generate and measure the quasi-2D flow as well as the characterization of the generated flow. Section \ref{Sec_ActiveMatter} introduces the animal used in this work called \textit{Artemia salina}.
A step by step protocol is provided for the experiments with \textit{A. salina} incorporated. The flows generated by swarming \textit{A. salina} without background flow are characterized in this section. Section \ref{Sec_theory} gives a brief theoretical background review for quasi-2D Navier-Stokes equation and Lagrangian coherent structures. The influence of swarming active matter on LCSs are then presented and discussed in \S \ref{Sec_contrast}, \S \ref{Sec_PF} and \S \ref{Sec_extension}. Section \ref{Sec_contrast} contrasts the active-matter-incorporated flow with the background flow overlaying two different kinds of perturbations: random noise of the same energy input and spatiotemporally correlated perturbations of the same energy input. Through contrasting, we confirm the importance of two different kinds of interactions: one involves the interaction between the flows generated by individual active matter, the other involves the interaction between the active-matter-generated flow and the background flow environment. Section \ref{Sec_PF} probes further onto the question about how the effect of active matter on background flow changes with number density (packing fraction). Section \ref{Sec_extension}, on the other hand, explores how the effect changes with the variation of background flow intensity. Finally, we summarize our findings and draw conclusions in \S \ref{Sec_conclusion}.

\section{Experiment: two-dimensional electromagnetically driven flow}\label{2Dflow}
The quasi-2D flow as a background for examining the impact of swarming active matter on LCSs was generated by an electromagnetically driven flow system and resolved using a particle tracking velocimetry (PTV) system \cite{ouellette2006quantitative}. The experimental apparatus, particle tracking operations and some prime characteristics of the flow are introduced successively in this section.
\subsection{Apparatus}
\begin{figure}
  \centerline{\includegraphics{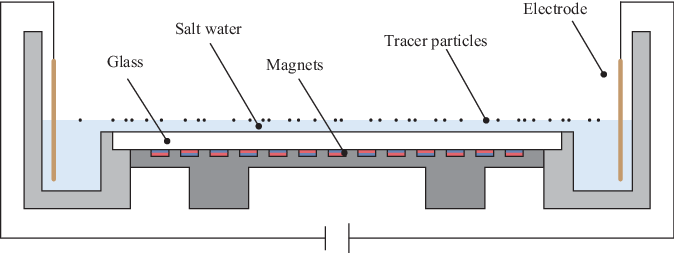}}
  \caption{Schematic diagram of the experimental apparatus.}
\label{Fig_setup}
\end{figure}

The apparatus for the quasi-2D flow system was composed of a main acrylic body frame, a pair of copper electrodes installed on the opposite sides of the setup and a piece of tempered glass in the center that separated a thin layer of salt water on top and an array of cylindrical magnets below. A schematic diagram for the setup is shown in Fig. \ref{Fig_setup}. Similar apparatus has been introduced in detail elsewhere \cite{kelley2011onset,si2022preferential,fang2018influence}, hence here we only briefly describe the necessary parameters.
The dimension of the main body frame and the glass floor in the center were 96.5 $\times$ 83.8 cm$^2$ and 81.3 $\times$ 81.3 cm$^2$, respectively. The upper surface of the glass was coated with hydrophobic materials (Rain-X) to reduce friction and the lower surface was covered by light-absorbing blackout film. Beneath the glass, the cylindrical magnet array was organized in checkerboard pattern of alternative polarity with a center-to center space of $L_{m} = $ 5.1 cm. Each magnet (neodymium grade N52) had an outer diameter of 1.27 cm and a thickness of 0.64 cm, with the maximum magnetic flux density of 1.5 T at the magnet surface. We loaded a thin layer (6 mm thickness) of 14$\%$ by mass NaCl solution on top of the glass. The solution had density $\rho =$ 1.101g/cm$^3$ and viscosity $\nu = 1.25 \times 10^{-2}$ cm$^2$/s. By passing a dc current through the conducting solution layer, we were able to drive a quasi-2D flow with the resulting Lorentz body force in horizontal directions and control the flow Reynolds number by adjusting the dc current intensity. The 2D was well kept throughout our experiments. Note that under our definition of $Re$ (see \S \ref{SubSec_flowChar}), since the length scale $L_m$ was arbitrary, the $Re$ was just a characterization of the flow velocity.

\subsection{Particle tracking and data post-processing}
To track the flow, we seeded the fluid with green fluorescent polyethylene particles (Cospheric) with a density of 1.025 g/cm$^3$ and diameters ranging from 106 to 125 $\mu$m. The Stokes number of the particles was of order 10$^{-3}$, which means that the particle could accurately trace the flow \cite{ouellette2008transport}. Since the density of the particles was lower than the working fluid, they would float on the gas-liquid interface and show a slow clustering tendency due to surface tension effects, which is known as the ``cheerios effect" \cite{vella2005cheerios}. To reduce the surface tension, a small amount of surfactant was added to the fluid in order to minimize its impact on the movement of the tracers. 

We used a machine vision camera (Basler, acA2040-90$\mu$m) to image the flow that was illuminated by blue LED lights. A 19.4 cm by 19.4 cm region at the center of the setup was recorded with a resolution of 1600 pixels by 1600 pixels. About 8000 particles could be recorded at a frame rate of 60 frames per second. With this particle density and frame rate, we could obtain a highly spatiotemporally resolved velocity field through a particle tracking velocimetry (PTV) algorithm. For easier use, the measured flow was then interpolated onto regular Eulerian grids using cubic interpolation with a grid size of 15 pixels (1.8 mm), which gave a grid density similar to the original particle density. 

When the images included both particles and \textit{A. salina}, the pixels corresponding to \textit{A. salina} were pre-masked before the images were processed by the PTV algorithm. The intensity of these pixels were replaced by the intensity value of the black background so that they would not be detected by the PTV algorithm. Therefore, only the motion of the flow would be tracked.

\subsection{Flow characterization}\label{SubSec_flowChar}
\begin{figure}
  \centerline{\includegraphics{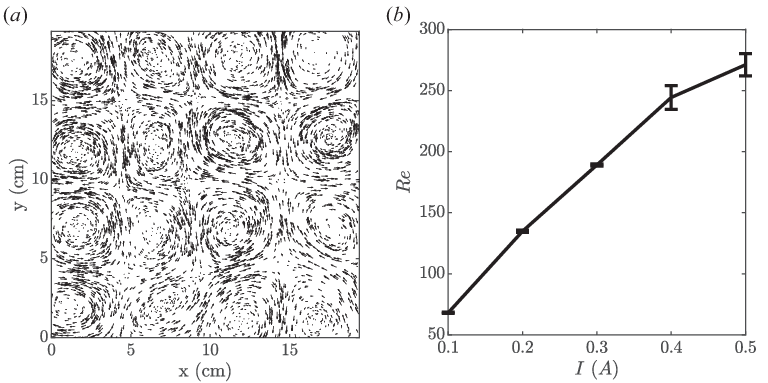}}
  \caption{(\textit{a}) A snapshot of the measured flow field under 0.3 A dc current. Velocity vectors are downsampled for better visualization. (\textit{b}) The mean $Re$ over 5000 frames of the measured flow fields under dc current intensities from 0.1 A to 0.5 A. The error bar represents one standard deviation.}
\label{Fig_FlowCharacterization}
\end{figure}

In this work, we measured flows under dc current intensities from 0.1 A to 0.5 A with a 0.1 A increment. Under 0.1 A, the flow was a steady laminar cellular flow. As the current intensity increased, the flow gradually became weakly time-dependent but was still strongly correlated with the locations of the magnets, where the strongest forcing was located. Figure \ref{Fig_FlowCharacterization}(a) shows a snapshot of the measured flow field under a 0.3 A dc current. It can be noticed that the center-to-center distance between flow cells was the same as $L_m$, which was used here as the characteristic length scale of the flow field. 
The Reynolds number calculated as $Re = UL_{m}/\nu$, where $U$ is the root-mean-square velocity of the flow field, increased from 68 to 270 with the corresponding dc current intensities (Fig. \ref{Fig_FlowCharacterization}(b)).

\section{Experiment: active matter}\label{Sec_ActiveMatter}
\begin{figure}
    \centering
    \centerline{\includegraphics{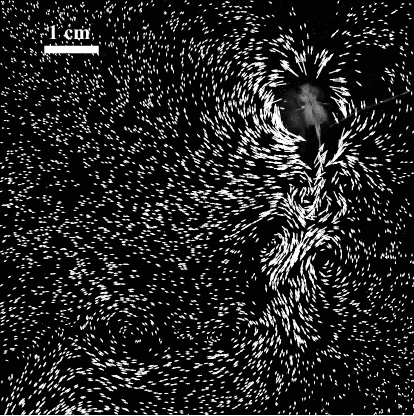}}
    \caption{An image of a single \textit{A. salina} swimming in quiescent water. The vectors mark the flow generated by this \textit{A. salina}.}
    \label{Fig_ShrimpJetMap}
\end{figure}

In this work, a typical centimeter-scale animal \textit{Artemia salina}, commonly known as brine shrimp, was used to examine the influence of active matter on LCSs. The typical body length of an adult-sized \textit{A. salina} is about 1 cm \cite{ouillon2020active}. This small animal swims by propelling its pairs of appendages and forms a jet opposite to its swimming direction. Figure \ref{Fig_ShrimpJetMap} shows a snapshot of the flow field around a single \textit{A. salina} swimming in quiescent water. It can be noticed that an \textit{A. salina} can generate flows at length scales much larger than its body length, which creates chances for dynamical interactions with background flow at larger scales. In this section, we will describe in detail the protocol for the experiments with \textit{A. salina} incorporated into the quasi-2D flow. The flows generated by swarming \textit{A. salina} at different packing fractions (PF) are then characterized in \S \ref{SubSec_SwarmFlow}.

\subsection{Protocol for experiments with animal}\label{SubSec_Protocol}
Adult-sized \textit{A. salina} was obtained from the supplier (Northeast Brine Shrimp) and cultured in 3$\%$ by mass NaCl solution for at least 24 hours before they were introduced to the experiment. Since the culturing fluid was less dense than the working fluid, the \textit{A. salina} would be kept floating on the surface of the working fluid throughout the full experiment period.
Sets of experiments with different background flow intensities (dc current intensities) were conducted separately. For each set with the same background flow intensity and different packing fractions of \textit{A. salina}, the series of data were collected following the protocol as below:
\begin {enumerate}
    \item Load the working fluid to the apparatus.
    \item Turn on the dc power supply. Adjust the current intensity to the intended magnitude and wait until the flow is fully developed.
    \item Record the background flow with constant current intensity. (The root-mean-square velocity of the background flow for all sets is shown in Fig. \ref{Fig_FlowCharacterization}(b)).
    \item Turn off the dc power supply. Record the decaying flow with enough time length. This is used to calculate the bottom friction coefficient $\alpha$ \cite{fang2017multiple} (see \S \ref{SubSec_2DNS}).
    \item Introduce a suitable amount of \textit{A. salina} to the working fluid and wait until the animals spread out and reach a statistically stationary state. Record the flow with the introduced \textit{A. salina} with no background flow.
    \item Turn on the dc power supply. Adjust the current intensity to the intended magnitude and wait until the flow is fully developed. Record the flow with the introduced \textit{A. salina} and background flow.
    \item Turn off the dc power supply and let the flow decay.
    \item Repeat step (v) to (vii) by gradually introducing more \textit{A. salina} to the working fluid until a desired high density of \textit{A. salina} is reached.
\end{enumerate}

\subsection{Swarming-\textit{A. salina}-induced flow characterization}\label{SubSec_SwarmFlow}
\begin{figure}
  \centerline{\includegraphics{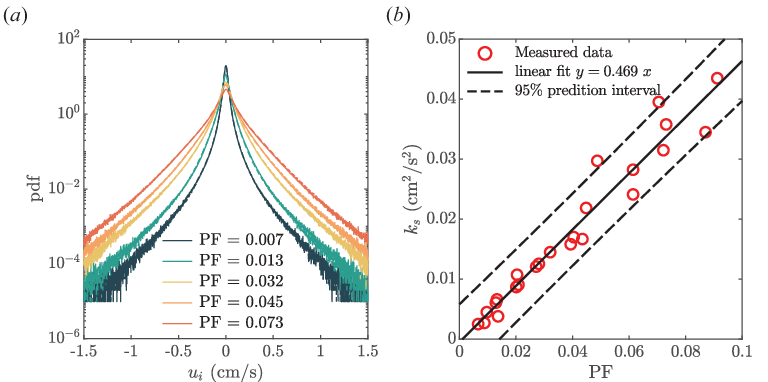}}
  \caption{(\textit{a}) Velocity component magnitude distribution for the flow field with different selected packing fractions (PF) of \textit{A. salina} and no background flow. The $u$ components are plotted in solid lines and the $v$ components are plotted in dashed lines. However, since the two lines are highly overlapped, the dashed line becomes less discernible. This means that both components with the same PF show the same distribution, which implies that the movements of \textit{A. salina} have no directional bias. (\textit{b}) Turbulent kinetic energy $k_s$ of flow generated by swarming \textit{A. salina} at different PF. Red circles represent measured data for all sets. The solid line shows the least squares regression of the data and the dashed line gives the 95$\%$ prediction interval (two standard deviations).}
\label{Fig_ShrimpFlowCharacterization}
\end{figure}

Since the light environment for the experiment was uniformly deployed, the \textit{A. salina} had no directional bias and showed a swarming movement pattern. As is shown in Fig. \ref{Fig_ShrimpFlowCharacterization}(a), the $u$ component and $v$ component of the flow field show exactly the same distribution for all selected packing fractions, where $u$ and $v$ are velocity components of two perpendicular axes of the 2D flow field. The flow field induced by swarming \textit{A. salina} was strongly non-Gaussian with a heavy long tail, indicating the occurrence of remarkable rare events. 
We also examined how the turbulent kinetic energy of the flow would increase with the packing fraction (PF) of \textit{A. salina}. Here, in two dimensions, PF is defined as the ratio of the area occupied by \textit{A. salina} to the total area of the image domain. Therefore, PF serves as a direct measure of the number density of \textit{A. salina} in the flow field.
In Fig. \ref{Fig_ShrimpFlowCharacterization}(b), we show the data measured in step (v) of \S \ref{SubSec_Protocol} for all sets of experiments (since step (v) was conducted without dc current, the experimental condition could be considered as the same for all sets of experiments at this step) together with the list squares regression.
We found that the swarming-\textit{A. salina}-induced turbulent kinetic energy $k_s = \frac{1}{2}\left<\boldsymbol{u'\cdot u'}\right>$ increased linearly with the packing fraction of \textit{A. salina}, where the $\boldsymbol{u'}$ is the fluctuation velocity and here, due to the swarming movement, $\boldsymbol{u'}$ equals to the full velocity $\boldsymbol{u}$. This indicates that the flow kinetic energy induced by individual \textit{A. salina} can be linearly added together to approximate the total kinetic energy yielded by swarming groups. Consequently, the turbulent kinetic energy at any packing fraction of \textit{A. salina} between 0 and 0.1 can be estimated by interpolation.

\section{Theoretical background}\label{Sec_theory}
\subsection{Two-dimensional Navier-Stokes equation}\label{SubSec_2DNS}
The equation of motion for incompressible flow in two dimensions is \cite{boffetta2012two}:
\begin{equation}\label{Eqn_2DNS}
    \partial_t{\boldsymbol{u}} + \boldsymbol{u}\nabla \boldsymbol{u} = \nabla p/\rho + \nu\Delta\boldsymbol{u} + \boldsymbol{f} + \alpha\boldsymbol{u},
\end{equation}
where $\boldsymbol{u}$ is the velocity, $p$ is the pressure and $\boldsymbol{f}$ is an external force term that drives the flow. Specifically, $\alpha$ is the coefficient corresponding to linear damping. Since in 2D flow, kinetic energy cascades toward larger scales rather than small scales \cite{kraichnan1967inertial}, the linear damping term dominates the energy dissipation process in 2D, as opposed to the viscous dissipation term in three-dimensions (3D).
In our experiment, the linear damping mainly results from the bottom friction. 
Without external forces and only focusing on the kinetic energy contained in the flow system under linear damping,
we get $\partial_t{\boldsymbol{u}} = \alpha\boldsymbol{u}$ ; and we will therefore, expect an exponential decay of the kinetic energy. The linear damping coefficient $\alpha$ can then me estimated by measuring a decaying flow \cite{fang2017multiple}. In this work, $\alpha$ values were measured at the start of each set of experiments (see \S\ref{SubSec_Protocol} step (iv)). The mean value of $\alpha$ was measured to be -0.074 s$^{-1}$, with one standard deviation at 0.009 s$^{-1}$.

\subsection{Lagrangian coherent structures and finite-time Lyapunov exponent}\label{LCS}
Flows show coherence ubiquitously. 
The coherent structures - macroscopic regions in the flow that show distinguishable spatial and temporal correlations - are hence significant in characterizing complex flows in a simpler lower-order way. 
Compared to Eulerian coherent structures, such as vortices, detecting coherent structures in a Lagrangian way, that is, following the paths of tracers in the flow, inherently shows many advantages because it accounts for the effects of advection.
Lagrangian coherent structures (LCSs), as described by \cite{haller2015lagrangian}, are regions in the flow that mark the most repelling, attracting, and shearing material surfaces. Since the LCSs detect the most straining regions in the flow, they are considered as effective representations of transport barriers in the flow that is crucial for transportation and mixing processes. 
In 2D, due to the inverse energy cascade, the LCSs are more robust than in 3D, which makes the 2D flow a good test system for investigating LCSs dynamics.

One of the most commonly used methods for detecting LCSs is the finite-time Lyapunov exponent (FTLE). In two dimensions, we calculate FTLE following the method in \cite{haller2015lagrangian}. The flow gradient $\nabla F^{t}_{t_0}(\boldsymbol{x_0})$ is first calculated using a finite-difference approximation as
\begin{equation}
    \nabla F^{t}_{t_0}(\boldsymbol{x_0}) \approx 
    \begin{pmatrix}
    \frac{x(t;t_0,\boldsymbol{x_0}+\boldsymbol{\delta_1}) - x(t;t_0,\boldsymbol{x_0}-\boldsymbol{\delta_1})}{|2\boldsymbol{\delta_1}|} & \frac{x(t;t_0,\boldsymbol{x_0}+\boldsymbol{\delta_2}) - x(t;t_0,\boldsymbol{x_0}-\boldsymbol{\delta_2})}{|2\boldsymbol{\delta_2}|}\\
    \frac{y(t;t_0,\boldsymbol{x_0}+\boldsymbol{\delta_1}) - y(t;t_0,\boldsymbol{x_0}-\boldsymbol{\delta_1})}{|2\boldsymbol{\delta_1}|} & \frac{y(t;t_0,\boldsymbol{x_0}+\boldsymbol{\delta_2}) - y(t;t_0,\boldsymbol{x_0}-\boldsymbol{\delta_2})}{|2\boldsymbol{\delta_2}|}
    \end{pmatrix}
    ,
\end{equation}
where $F^{t}_{t_0}(\boldsymbol{x_0})$ = $\boldsymbol{x}(t;t_0,\boldsymbol{x_0})$ is the flow map that depicts the transport of a fluid element from an initial position $\boldsymbol{x_0}$ at time $t_0$ to the position $\boldsymbol{x}$ at time $t$. The $\boldsymbol{\delta_1}$ and $\boldsymbol{\delta_2}$ are two vectors pointing in horizontal ($x$) and vertical ($y$) directions, respectively, that represent small perturbations.
The right Cauchy-Green strain tensor
\begin{equation}
    C(\boldsymbol{x_0}) = [\nabla F^{t}_{t_0}(\boldsymbol{x_0})]^T\nabla F^{t}_{t_0}(\boldsymbol{x_0})
\end{equation}
depicts how an initially small perturbation at $\boldsymbol{x_0}$ evolves with time. The eigenvalues $\lambda_i (\boldsymbol{x_0})$ and eigenvectors $\xi_i(\boldsymbol{x_0})$ of this symmetric and positive definite tensor satisfies
\begin{equation}
    C\xi_i = \lambda_i\xi_i,\mbox{ }i = 1,2;\mbox{ } 0<\lambda_1<1<\lambda_2,\mbox{ }\xi_1\perp\xi_2.
\end{equation}
The FTLE is then calculated as
\begin{equation}
    \text{FTLE}(\boldsymbol{x_0}) = \frac{1}{t-t_0}log\sqrt{\lambda_2(\boldsymbol{x_0})}.
\end{equation}
Repelling FTLE (FTLE$^+$) is calculated by integrating in forward time ($t_0 < t$). The regions with local maximum values usually occur along ridges and the ridges are interpreted as effective detection of repelling LCSs. Attracting FTLE (FTLE$^-$), on the other hand, is calculated by integrating in backward time ($t_0 > t$). Since FTLE$^-$ is negative, regions with local minimum values then form ridges that are the locations of attracting LCSs \cite{si2021preferential}.

\begin{figure}
    \centering
    \centerline{\includegraphics{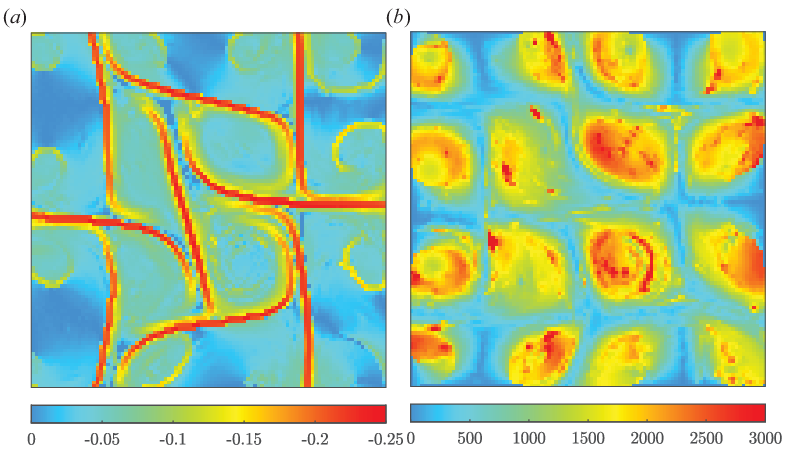}}
    \caption{(\textit{a}) Snapshot of one frame of attracting FTLE calculated with the flow under 0.3 A dc current. The flow was integrated by two eddy turnover time. The unit of the FTLE is s$^{-1}$.
    (\textit{b}) Snapshot of the LAVD for the same flow field in (\textit{a}) at the same moment. The flow was integrated by one eddy turnover time.}
    \label{Fig_FTLE_demo}
\end{figure}

In this work, we took the magnitude of both $\boldsymbol{\delta_1}$ and $\boldsymbol{\delta_2}$ to be 15 pixels (1.8 mm). The flow was integrated by two eddy turnover time ($L_m/U$) for each set of experiment. Figure \ref{Fig_FTLE_demo}(a) shows a snapshot of attracting FTLE (FTLE$^-$) for the flow under 0.3 A dc current intensity. The FTLE$^-$ ridges mark the most straining hyperbolic regions that separate the elliptical islands.

Additionally, \cite{haller2016defining} also introduced Lagrangian-averaged vorticity deviation (LAVD) to detect rotationally coherent vortices. In 2D, the LAVD is calculated as:
\begin{equation}
    \text{LAVD}^{t}_{t_0}(\boldsymbol{x_0}) = \int^{t}_{t_0}|\omega(\boldsymbol{x}(s;\boldsymbol{x_0}),s) - \overline{\omega}(s)|ds,
\end{equation}
where $\omega$ is the vorticity and $\overline{\omega}$ is the instantaneous spatial mean of $\omega$. Figure \ref{Fig_FTLE_demo}(b) displays a snapshot of the LAVD field for the same flow field in Fig. \ref{Fig_FTLE_demo}(a) at the same moment. The flow was integrated by one eddy turnover time. 
By examining the FTLE field and LAVD field of a flow, we can characterize the straining hyperbolic regions and the rotating elliptical regions and explore how these regions change with additional external forces.

\section{The impact of active matter on LCSs - describe by contrasting}\label{Sec_contrast}
The incorporation of active matter into an existing flow field introduces extra energy, new boundary conditions, as well as new time and length scales. The profound complexity of the system makes a clear understanding of the full dynamical process extremely challenging. 
Therefore, in this section, we choose to first describe the influence of active matter on LCSs by contrasting it with the influence of two other kinds of perturbations: overlaid random noise of the same kinetic energy and overlaid spatiotemporally correlated perturbations of the same kinetic energy. The hope is that, by contrasting with the influence of other kinds of perturbations, we can gain insights into the impact of active matter on LCSs. While it is challenging to fully comprehend what the influence ``is", we can at least acquire some knowledge about what it ``is not". In this and the next section, we focus on the set of data measured with $I$ = 0.3 A. Specifically, from now on, we call the flow that actually embedded the active matter ``active-matter-incorporated flow'' contrasting to the cases where the flow's Eulerian velocity fields were overlaid with perturbation velocity fields. In addition, the LCSs in our flow had a length scale of roughly 10 cm, and the \textit{A. salina} had a length scale of 1 cm.

\begin{figure}
    \centering
    \centerline{\includegraphics{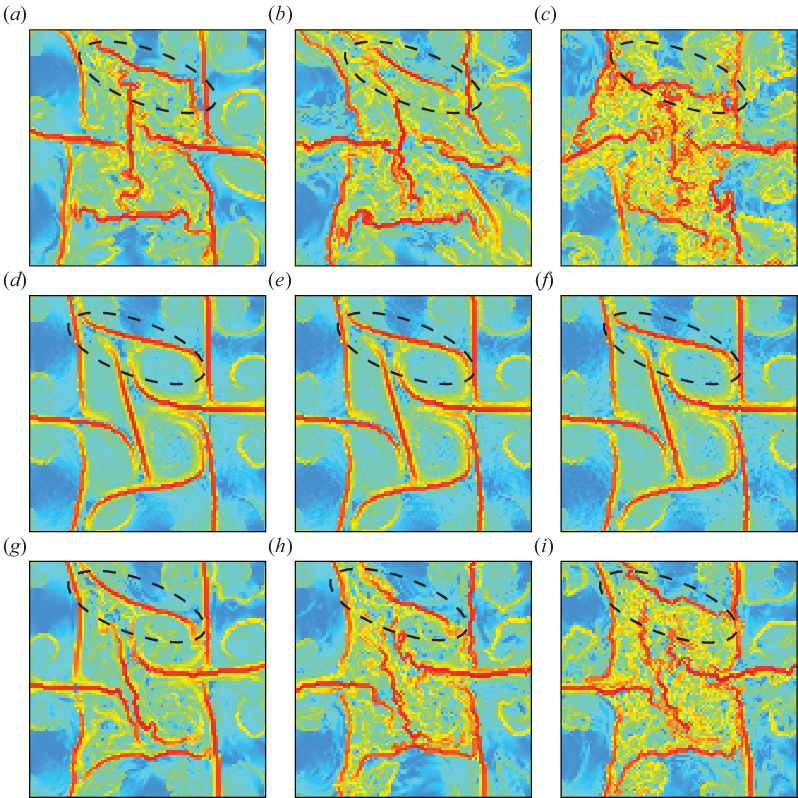}}
    \caption{(\textit{a})-(\textit{c}) give snapshots of FTLE$^-$ fields with \textit{A. salina} incorporated into the background flow. The packing fractions of \textit{A. salina} are (\textit{a}) 0.014, (\textit{b}) 0.029, and (\textit{c}) 0.074. The corresponding $k_s/\bar{E}$ are 0.076, 0.15, and 0.39, respectively.
    (\textit{d})-(\textit{f}) give snapshots of FTLE$^-$ fields with local random noise overlaid to the background flow. The packing fractions selected for obtaining $k_s$ are (\textit{d}) 0.02, (\textit{e}) 0.04, and (\textit{f}) 0.08. The corresponding $k_s/\bar{E}$ are 0.086, 0.18, and 0.36, respectively.
    (\textit{g})-(\textit{i}) give snapshots of FTLE$^-$ fields with \textit{A. salina} generated flow in quiescent water overlaid to the background flow. The packing fractions of \textit{A. salina} are (\textit{g}) 0.013, (\textit{h}) 0.032, and (\textit{i}) 0.073. The corresponding $k_s/\bar{E}$ are 0.063, 0.14, and 0.36, respectively.
    The incorporation cases yield stronger shifting and twisting effects on LCSs.
    The reader can use the FTLE$^-$ ridges in the circled region to compare the differences.
    The colorbar range is the same as Fig. \ref{Fig_FTLE_demo}(a).
    }
    \label{Fig_FTLE_compare}
\end{figure}

A good starting point is to compare the impacts generated by perturbations from \textit{A. salina} with those from local random noise. One of the most significant differences between these two conditions is that the perturbations from \textit{A. salina} have spatiotemporal correlations. As is mentioned in \S \ref{Sec_ActiveMatter}, the perturbation generated by a single \textit{A. salina} is in the form of a jet opposite to its swimming direction. The jets generated by swimming \textit{A. salina} interact with each other and also interact with the background flow. These interaction processes persist in time and could form large scale flow structures, introducing extra time and length scales.
Local random noise, on the other hand, doesn't have any spatial or temporal correlations. By comparing these two conditions, we can reveal the importance of the new time and length scales introduced by active matter.
The local random noise was generated with random directions and constant magnitude.
In Fig. \ref{Fig_ShrimpFlowCharacterization}(b), we characterized the linear relationship between the packing fraction of \textit{A. salina} and the turbulence kinetic energy $k_s$ generated by \textit{A. salina}. Therefore, the magnitude of the random noise corresponding to a certain packing fraction could be acquired through the list squares regression line in Fig. \ref{Fig_ShrimpFlowCharacterization}(b). Next, we overlaid the generated random noise to the measured background flow field (the background flow field was measured in step (iii) of \S \ref{SubSec_Protocol}). We then integrated the resulting flow field to obtain the FTLE$^-$ fields. Figure \ref{Fig_FTLE_compare}(d)-(e) shows the snapshots of FTLE$^-$ fields at the same moment with increasing packing fractions at 0.02, 0.04, and 0.08. The corresponding $k_s$ normalized by the mean flow energy $\bar{E} = \frac{1}{2} \left<\boldsymbol{u}\right>\cdot\left<\boldsymbol{u}\right>$ for the background flow field were 0.086, 0.18, and 0.36, respectively. The increase in perturbation intensity created a diffusion-like effect that made the FTLE$^-$ ridges slightly smear out. However, the essential structure of the LCSs was not affected at all. 
On the opposite, incorporating \textit{A. salina} into the flow yielded remarkably contrasting results.
Figure \ref{Fig_FTLE_compare}(a)-(c) shows the snapshots of FTLE$^-$ fields with measured packing fractions of \textit{A. salina} at 0.014, 0.029, and 0.074. The ratio of the turbulence kinetic energy $k_s$ to the mean flow energy $\bar{E}$ were 0.076, 0.15, and 0.39, respectively. With the increase of number density of \textit{A. salina}, the LCSs were significantly distorted. The FTLE$^-$ ridges exhibited a more intricate and twisting configuration. The large LCSs structures also diverged intensively from the original cellular form. From this contrasting, it shows clearly that the perturbations induced by active matter are completely different from local random noise of similar energy. With the same intensity of $k_s$, active matter yielded a much more dramatic influence on twisting and deforming the LCSs than local random noise. The spatial and temporal correlation involved in the active-matter-induced perturbation played a significant role in altering the LCSs of the flow.

The comparison with local random noise demonstrates the important role played by the spatiotemporal correlations in the perturbations created by active matter. This spatiotemporal correlation results from the interactions between large scale flow structures.
However, considering the condition where the movements of active matter are incorporated into the background flow, the interaction not only occurs among active-matter-induced flows, but also occurs between the active-matter-induced flows and the background flow field. Through the next contrasting between the active-matter-incorporated flow and the background flow directly overlaying the active-matter-induced flow in quiescent water, we will show that the interaction between the active-matter-induced flow and the background flow is also a non-negligible factor on shifting and twisting LCSs. 

\begin{figure}
    \centering
    \centerline{\includegraphics{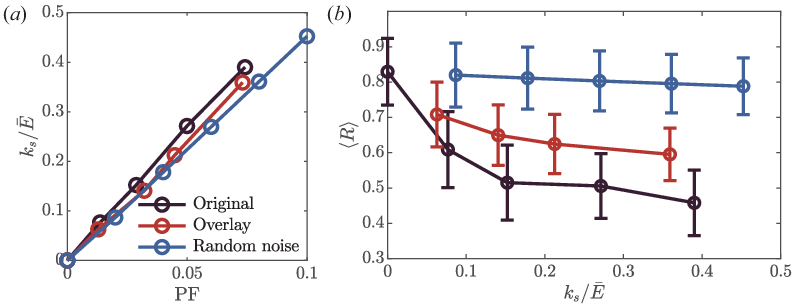}}
    \caption{(\textit{a}) The ratio of the swarming-active-matter-induced turbulent kinetic energy $k_s$ to the mean kinetic energy of the background flow $\bar{E}$ at different packing fractions (PF) for the original active-matter-incorporated flow (black line), the background flow overlaying the active-matter-induced flow in quiescent water (red color), and the background flow overlaying local random noise. (\textit{b}) The mean correlation coefficient $\left<R\right>$ of FTLE$^-$ fields at different kinetic energy ratio $k_s/\bar{E}$. The mean is calculated for all possible pairs of FTLE$^-$ fields within all one eddy turnover time (7.5 s) windows. The FTLE$^-$ fields have 100 frames in total with 0.5 s time interval. The error bar shows one standard deviation. Different colors represent the same condition as (\textit{a}).}
    \label{Fig_FTLE_corr}
\end{figure}

We used the flow fields measured at step (v) in \S \ref{SubSec_Protocol} and directly overlaid them onto the background flow field measured at step (iii). The overlaid flow field included the information from both the background flow and the swarming-active-matter-induced flow. Actually, at the same packing fraction, the turbulence kinetic energy was similar no matter whether the active matter was overlaid or incorporated (Fig. \ref{Fig_FTLE_corr}(a)). The subtle difference here is that the overlaid case does not have ``real" dynamics about the interaction between active-matter-induced flows and the background flow. Considering the dynamics in a spectral view may make the difference between these two cases clearer. The background flow had an energy injection length scale $L_m$ and the kinetic energy shall be mainly stored in large scale eddies having length scales greater than $L_m$ because the quasi-2D flow system had a net inverse energy flux. The flow induced by \textit{A. salina}, however, had a much smaller energy injection length scale than the background flow. Therefore, the flow measured for swarming \textit{A. salina} in quiescent water shall have main dynamics at smaller length scales. The overlaying of these two flow fields included the dynamics for both flow fields but was lack of the dynamics induced by the direct interaction between active-matter-induced flows and the background flow.
Through this contrast, we intend to isolate the dynamics that occurred at the length scales between the energy injection length scale of the background flow and the energy injection length scales of the swarming-active-matter-induced flow. The dynamics that occurred at these length scales were attributed to the interaction between the flow induced by the swarming active matter and the background flow field.

Figure \ref{Fig_FTLE_compare}(g)-(i) shows the snapshots of FTLE$^-$ fields at the same moment with measured packing fractions of \textit{A. salina} at 0.013, 0.032, and 0.073. The corresponding $k_s/\bar{E}$ ratio were 0.063, 0.14, and 0.36, respectively. It can be noticed by comparing the two overlaying cases that overlaying the swarming-active-matter-induced flow created a much more remarkable effect on deforming the LCSs than overlaying local random noise, which again confirms the importance of the spatiotemporal correlation involved in the swarming-active-matter-induced flow. But comparing the second overlaying case, the active-matter-incorporated flow cases (Fig. \ref{Fig_FTLE_compare}(a)-(c)) have more shifted and twisted LCSs. 
To further quantify the degree of deformation on LCSs, for each case, we calculated the FTLE$^-$ fields every 30 frames (0.5 s) for 3000 frames (50 s) of data and then quantified the mean correlation coefficient $\left<R\right>$ for all possible pairs of FTLE$^-$ fields within one-eddy-turnover-time windows in the same group of data. For example, the correlation coefficient $R_{A,B}$ between FTLE$^-$ field $A$ and FTLE$^-$ field $B$ was calculate as 
\begin{equation}
    R_{A,B} = \frac{\text{cov}(A,B)}{\sigma_A\sigma_B},
\end{equation}
where cov($A,B$) is the covariance of $A$ and $B$ and $\sigma_A$ and $\sigma_B$ are the standard deviation of $A$ and $B$. $\left<R\right>$ was calculated as the mean of all calculated $R_{A,B}$ values in the same group of data. The parameter $\left<R\right>$ quantifies the degree of deformation on LCSs as time progresses. Since the background flow was weakly time-dependent, when the time interval between frame A and frame B was large, the correlation became weak for all groups. Therefore, we let the time interval between A and B not be larger than one eddy turnover time.
As is shown in Fig. \ref{Fig_FTLE_corr}(b), incorporating swarming \textit{A. salina} yielded the strongest deformation. The local random noise, however, created a minor influence on LCSs with the increase of turbulence kinetic energy.

\begin{figure}
    \centering
    \centerline{\includegraphics{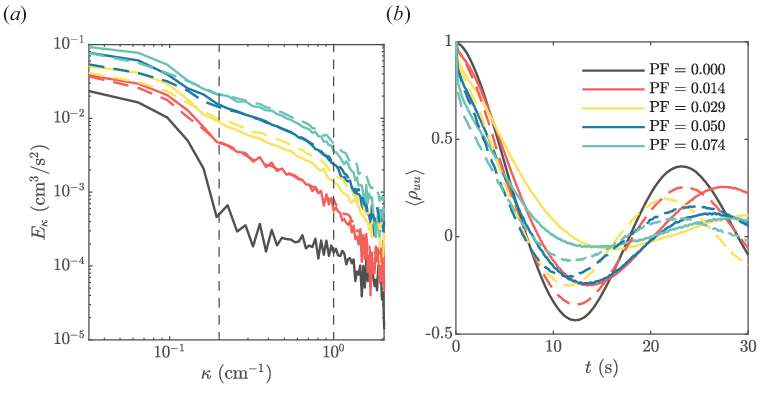}}
    \caption{(\textit{a}) The turbulence kinetic energy spectrum for the \textit{A. salina}-incorporated flow (solid line) and the background flow overlaying the swarming-\textit{A. salina}-induced flow in quiescent water (dashed line). The packing fraction for each case is the same as (\textit{b}). The left vertical dash line marks the energy injection length scale of the quasi-2D flow system $L_m$. The right vertical dash line marks the length scale of a single \textit{A. salina}.
    (\textit{b}) Lagrangian velocity correlation for the \textit{A. salina}-incorporated flow (solid line) and the overlaying flow (dashed line).}
    \label{Fig_scale_analysis}
\end{figure}

Figure \ref{Fig_FTLE_corr}(b) reveals the importance of two different dynamical interactions. The first interaction involves the interaction between the flows induced by individual active matter. The second interaction involves the interaction between the swarming-active-matter-induced flow and the background flow field. Both interactions play a non-negligible role in influencing the formation of LCSs. To further illustrate the second kind of interaction, we performed spectral analysis and Lagrangian statistics for both the active-matter-incorporated flow and the background flow overlaying swarming-\textit{A. salina}-induced flow. As was discussed above, pure swarming-active-matter-induced flow in quiescent water exhibited a smaller characteristic length scale than the background flow. The interaction between these two flows, however, could yield flow structures on scales larger than the pure swarming flow. From Fig. \ref{Fig_FTLE_corr}(a), we see the turbulence kinetic energy $k_s$ was the same for both cases. This indicates different spectral distributions of $k_s$ for these two cases. The active-matter-incorporated flow should have a higher portion of $k_s$ stored at larger scales than the overlaid case. This inference was confirmed by Fig. \ref{Fig_scale_analysis}(a). The case that incorporated \textit{A. salina} did exhibit higher spectral energy than the overlaid case at lower wave numbers $\kappa$ and lower spectral energy than the overlaid case at higher wave numbers. This difference was not salient, though, because of the limited amount of energy introduced by the active matter.
However, the redistribution of energy is consistently observed in all packing fractions. The small amount of energy redistribution notwithstanding, it is surprising that such a subtle difference in power spectrum could result in such pronounced differences in LCSs structures (Fig. \ref{Fig_FTLE_corr}(b)). This result shows that even a small amount of active matter could lead to a profound impact on transport and mixing. 

Figure \ref{Fig_scale_analysis}(b) essentially reveals the same physics as Fig. \ref{Fig_scale_analysis}(a). In this figure, we show the mean Lagrangian velocity correlation $\left<\rho_{uu}\right>$ of the velocity component $u$. For each tracer following a Lagrangian trajectory as $\boldsymbol{x}(t_;t_0,\boldsymbol{x_0})$, $\rho_{uu}$ was calculated as \cite{pope2000turbulent}
\begin{equation}
    \rho_{uu}(t) = u(\boldsymbol{x}(t;t_0,\boldsymbol{x_0}))u(\boldsymbol{x}(t_0;t_0,\boldsymbol{x_0})).
\end{equation}
Compared with the overlaid case, $\left<\rho_{uu}\right>$ decreased slower in the active-matter-incorporated flow at the initial stage but displayed a weaker periodicity over longer distances. The $\left<\rho_{uu}\right>$ decreased slower at the initial stage means the tracers had a longer ``memory", which indicates the stronger correlation at larger length scales, which results from the interaction between the active-matter-induced flow and the background flow. On the other hand, the weaker periodicity over long distances indicates a more intensive deformation and distortion on the LCSs, which breaks the initial cellular flow structure.

\section{The impact of active matter on LCSs - effect of different packing fractions}\label{Sec_PF}
In the previous section, we contrasted the \textit{A. salina}-incorporated flow with the background flow overlaying both random noise and swarming-\textit{A. salina}-induced flow field in quiescent water. Through contrast, we revealed the importance of two kinds of interactions in the active-matter-incorporated flow, which resulted in a stronger deformation effect on LCSs than the other two overlaying cases. In this section, we probe deeper into the active-matter-incorporated flow and examine how the effect of active matter on background flow changes with their packing fractions.

The background cellular flow can be categorized into two different regions: the hyperbolic regions and the elliptical regions. The hyperbolic regions exhibit the strongest stretching and compressing, which consequently have the strongest mixing \cite{balasuriya2016hyperbolic}. These regions can be identified by hyperbolic LCSs.
The elliptical regions, on the other hand, are dominated by rotational movement. If we use the FTLE$^-$ field to identify these two regions, the FTLE ridges with the lowest values should mark the hyperbolic LCSs. The elliptical regions, due to their rotational movement, should have higher FTLE$^-$ values. Note that FTLE$^-$ values are negative. Thus, lower values mean stronger straining.
\begin{figure}
    \centering
    \centerline{\includegraphics{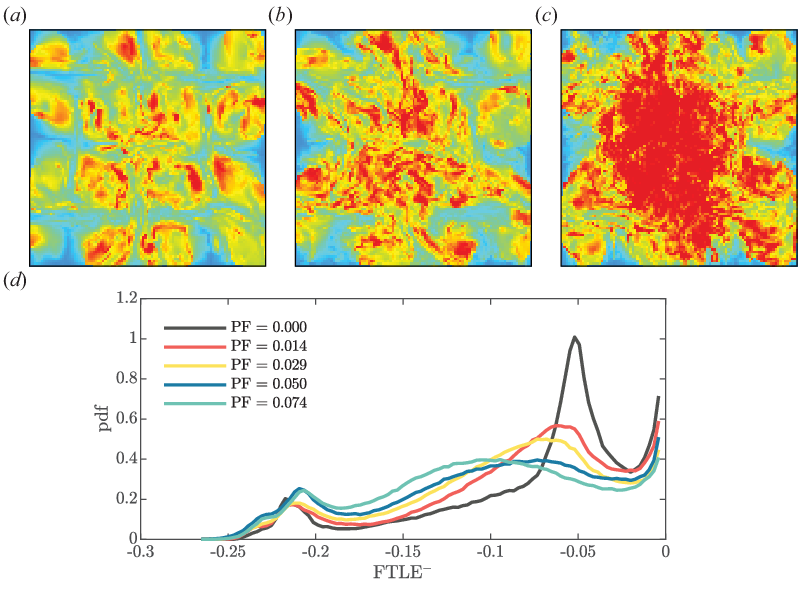}}
    \caption{(\textit{a})-(\textit{c}) gives a snapshot of the LAVD field for the same flow as Fig. \ref{Fig_FTLE_corr}(\textit{a})-(\textit{c}) at the same moment. The flow was integrated by 7.5 s, which was one eddy turnover time. The colorbar range is the same as Fig. \ref{Fig_FTLE_demo}(b). (d) gives the pdf of FTLE$^-$ values for \textit{A. salina}-incorporated flow with different packing fractions (PF).}
    \label{Fig_lavd_compare}
\end{figure}

To examine the impact of swarming \textit{A. salina} on the flow, we calculated the probability density function (PDF) of the FTLE$^-$ fields at different packing fractions of \textit{A. salina}. As is shown in Fig. \ref{Fig_lavd_compare}(d), for the background flow (black line) without \textit{A. salina}, there are two distinct peaks. The higher peak with larger FTLE$^-$ values represents the elliptical regions that occupy most of the regions in the flow. The lower peak, characterized by smaller FTLE$^-$ values, represents the hyperbolic regions. As the packing fraction of \textit{A. salina} increases, the peak corresponding to the elliptic regions drops substantially and spreads out simultaneously. The FTLE$^-$ value corresponding to the hyperbolic regions also moves to a lower value but in a much weaker extent than the elliptical region. This change indicates that the elliptical regions were substantially disturbed by the movement of \textit{A. salina}, even at a very small packing fraction, while the hyperbolic structures are more robust against perturbations from active matter. To further confirm the change in the elliptical region, we calculated the LAVD field for the \textit{A. salina}-incorporated flow. Figure \ref{Fig_lavd_compare}(a)-(c) show the snapshots of the LAVD fields for the same flow field in Fig. \ref{Fig_FTLE_compare}(a)-(c) at the same moment. It can be noticed that the elliptical islands were broken at even low packing fractions of \textit{A. salina}.
The peak corresponding to the hyperbolic regions, on the other hand, does not exhibit a change as strong as the elliptic regions. It shifts slightly toward a higher FTLE$^-$ value and has small changes in peak value. This does not mean, however, the hyperbolic regions were not affected by the movement of \textit{A. salina}. Since the hyperbolic regions were more robust than the elliptic regions, they were deformed and twisted by the \textit{A. salina}-induced fluctuations, as was demonstrated by FTLE ridges in Fig. \ref{Fig_FTLE_corr}(a)-(c), rather than be totally broken as the elliptical regions.

Therefore, a relatively more comprehensive picture of how swarming active matter affects the flow field with the increase of packing fraction can be depicted as follows: under low packing fractions, the rotational elliptic regions are first disturbed; regions within each elliptical island acquire a more intensive mixing effect. The hyperbolic regions are not significantly affected under low packing fractions of active matter and hence still work as transport barriers between elliptical islands. As the packing fraction of active matter increases, the elliptical regions are totally broken. Fluid is intensively mixed up in elliptical regions. The hyperbolic regions are shifted and twisted due to the increasing fluctuation introduced and gradually become less efficient in working as transport barriers. The flow hence acquires a stronger mixing.

\section{The variation of background flow intensity - the impact of $k_s/\bar{E}$}\label{Sec_extension}
The preceding two sections illustrated how the active matter and their interactions with background flows played a significant role in impacting the deformation of LCSs and how the effect varied with the change of their packing fractions. 
The analysis in these sections was based on the data measured under $I = 0.3$ A. Under this background flow intensity, the maximum ratio between the turbulent kinetic energy induced by active matter $k_s$ and the mean kinetic energy of the background flow $\bar{E}$ was approximately 0.4. 
In this section, we continue the investigation by comparing the data measured under varied dc current intensities. The hope is to push the condition to a lower $k_s/\bar{E}$ ratio and identify active-matter-dominated and flow-dominated regimes based on $k_s/\bar{E}$. 

\begin{figure}
    \centering
    \centerline{\includegraphics{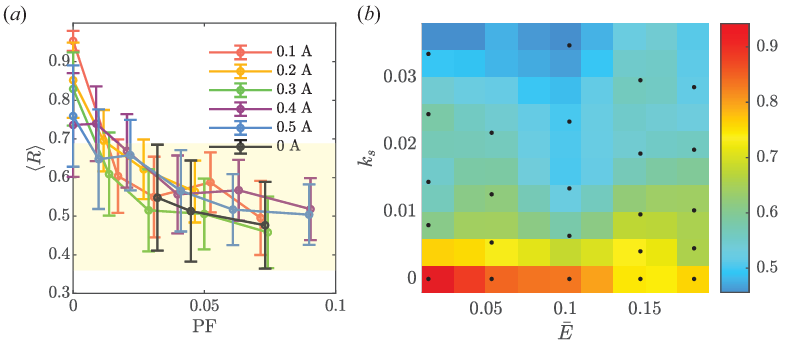}}
    \caption{(\textit{a}) The mean correlation coefficient $\left<R\right>$ of FTLE$^-$ fields under different background flow intensities. The legend shows the dc current intensities that drove the flows. The error bar represents one standard deviation. The shaded area marks the range of $\left<R\right>$ covered by pure active matter cases in quiescent water. (\textit{b}) gives the 2D map of $\left<R\right>$ interpolated from the data in (\textit{a}). The black points mark the measured data. The $k_s$ values were obtained through the least squares regression line in Fig. \ref{Fig_ShrimpFlowCharacterization}(b).}
    \label{Fig_FTLE_phase}
\end{figure}

We followed the method used in \S \ref{Sec_contrast} and calculated $\left<R\right>$ of FTLE$^-$ fields within one-eddy-turnover-time windows as the measure of LCSs deformation.
The results are presented in Fig. \ref{Fig_FTLE_phase}(a) and (b). Figure \ref{Fig_FTLE_phase}(a) shows directly all experimentally measured data points. These data points were interpolated onto a 2D map with $k_s$ and $\bar{E}$ being two axes in Fig. \ref{Fig_FTLE_phase}(b). We recall that $k_s$ and PF have a linear relationship for PF $<$ 0.1. The $k_s$ values in Fig. \ref{Fig_FTLE_phase}(b) were obtained through the least squares regression line in Fig. \ref{Fig_ShrimpFlowCharacterization}(b). 

There are two factors that can impact the $\left<R\right>$. The first is the impact of active matter. As we observe in Fig.\ref{Fig_FTLE_phase}(a) and (b), $\left<R\right>$ decreases as the packing fraction increases at all current intensities. Second, the flow becomes more and more time-dependent as the current increases. Even with no active matter (zero packing fraction), $\left<R\right>$ has a decreasing trend as dc current intensity increases. 

The cases with PF $\approx$ 0.02 ($k_s \approx 0.01$) are the smallest PF cases that we have besides the pure flow case (PF = 0). In these cases, we see a noticeable drop in $\left<R\right>$, and we already observe a decent amount of twisting in the LCSs (see Fig. \ref{Fig_FTLE_compare}). The minimum ratio of the measured $k_s/\bar{E}$ under this PF is around 5\% with 0.5 A dc current intensity and as indicated by the notable drop in $\left<R\right>$, we have already approached a visible change in LCSs. This result suggests that to reach the flow-dominated regime, the flow must contain very weak active matter, where the $k_s/\bar{E}$ ratio should be much smaller than 5\%.

To explore the active-matter-dominated regime, we calculated the $\left<R\right>$ for cases when the active matter was in quiescent fluid, where active matter dominated by definition. We did not consider the cases where the PFs were too small because the flow fields in those cases had a large portion of negligible velocity magnitudes that will contaminate the following calculation of the FTLE field. We observe that in all measured cases, as PF increases, $\left<R\right>$ approaches the value for the cases of pure active matter (the shaded area in Fig. \ref{Fig_FTLE_phase}(a)). The smallest ratio of $k_s/\bar{E}$ in the shaded area is around 10\%. That is the case with 0.5 A current ($\bar{E} \approx$ 0.2) and packing fraction of around 0.04 ($k_s \approx$ 0.02). Unobservantly, one may assume flow transition to the active-matter-dominated regime as $k_s/\bar{E}$ grows above 10\%. However, this conclusion may be flawed because the time dependence of the flow can also contribute to the lower $\left<R\right>$. We can imagine increasing the dc current to an extremely high value, in which case we can confidently say the external flow dominates the process. But the $\left<R\right>$ value will also drop simultaneously.
When we examine the change of $\left<R\right>$ as a function of PF, we will see that at PF = 0, $\left<R\right>$ is already low and close to the shaded area. We cannot conclude that the flow is active-matter dominated in this case. To eliminate the time dependency, we now focus on 0.2 A cases where the cellular structure strictly holds (We did not use 0.1 A cases because the smallest measured packing fraction had already entered the active-matter-dominated regime, and we could not determine the threshold for the active-matter-dominated regime). We observe that the smallest $k_s/\bar{E}$ ratio for $\left<R\right>$ to enter the shaded area is around 25\%. 

Because of the flow's time dependence as we increase the driving force, a clear separation between the flow-dominant regime and active-matter-dominant regime under very intensive and even turbulent background flow cannot be resolved in this work, and is pending further exploration. Nevertheless, we can make two conclusions. First, the incorporation of active matter, even at a very low $k_s/\bar{E}$ (as low as 5\% in our experiments), can yield a sharp drop of $\left<R\right>$ and the flow leaves the flow-dominated regime. Therefore, it would be expected that the flow-dominated regime exists when $k_s/\bar{E} \ll$ 5\%. Second, we reveal that the flow is active-matter-dominated when $\left<R\right> \ge 25$\%, as $\left<R\right>$, in this case, is indistinguishable from the $\left<R\right>$ of pure active matter case.

\section{Summary and conclusions}\label{Sec_conclusion}
We conducted experiments to study the interaction between swarming active matter and an external background flow field. In this work, we focused on the impact of swarming active matter on LCSs. 
\textit{A. salina}, a kind of centimeter-scale swimmer, was used as the proxy of active matter, and we introduced it into an electromagnetically driven quasi-2D cellular flow with LCSs length scale of around 10 cm. Both the number density of active matter and the dc current intensity were varied so that a broad range of $k_s/\bar{E}$ ratio was covered, where $k_s$ is the active-matter-induced turbulent kinetic energy and $\bar{E}$ is the mean kinetic energy of the background flow.

We illustrated the significance of two different kinds of interaction in impacting the deformation of LCSs by comparing the active-matter-incorporated flow with the background flow overlaying two different kinds of fluctuations. The first interaction involves the interaction between the flow produced by the agitations of individual active matter. The second interaction involves the interaction between the active-matter-induced flow and the background flow field. Both interaction processes play non-negligible roles in shifting and twisting the LCSs.

We further investigated the impact of active matter under different packing fractions. We found that the elliptical regions in the flow were first disturbed and gradually broken even at a small active matter packing fraction. The hyperbolic regions that worked as transport barriers of the flow were more robust and would not be totally broken by the agitations of active matter. However, this did not mean that hyperbolic regions were not affected. They exhibited strong deformation, characterized by strong shifting in positions and twisting in structures, as the packing fraction of active matter increased. This, by definition, resulted in a facilitated mixing effect.

Finally, we explored the effect of active matter on LCSs with varied background flow intensities. We found that the active matter could yield deformation of LCSs even at very low $k_s/\bar{E}$ ratios (5\%). The flow-dominated regime exists when $k_s/\bar{E} \ll 5$\%. In addition, we reveal that active matter starts to dominate when $k_s/\bar{E} \ge 25$\%. Our result suggests the possibility of breaking transport barriers and facilitating mixing through active matter in natural and industrial flows, such as forbidden zones in the coastal area \cite{olascoaga2006persistent}.  

Even though the dynamics of active matter have been extensively investigated from various perspectives, their interaction with an existing background flow field has long been overlooked. We hope that this work could provide insights into this problem and draw attention to the interaction between active matter and external flow structures.



\begin{acknowledgments}
This work was supported by the U.S. National Science Foundation under Grant No. CMMI-2143807.
\end{acknowledgments}






\bibliography{Biblio}

\end{document}